# Electrochemical Analysis of Na$_{0.74}$Co$_{1-x}$Nb$_x$O$_2$ ($x$ = 0, 0.05) as Cathode Materials in Sodium-ion Batteries

Jayashree Pati, Mahesh Chandra and Rajendra S. Dhaka[a)]

*Novel Materials and Interface Physics Laboratory,*
*Department of Physics, Indian Institute of Technology Delhi, Hauz Khas, New Delhi-110016 INDIA*

[a)]Corresponding author: rsdhaka@physics.iitd.ac.in

**Abstract.** Sodium-ion batteries (SIBs) have received significant attention as promising alternative for energy storage applications owing to the large availability and low cost of sodium. In this paper we study the electrochemical behavior of Na$_{0.74}$Co$_{1-x}$Nb$_x$O$_2$ ($x$ = 0 and 0.05 samples), synthesized via solid-state reaction. The Rietveld refinement of x-ray diffraction pattern reveals the hexagonal crystal symmetry with P63/mmc space group. The Na$_{0.74}$Co$_{0.95}$Nb$_{0.05}$O$_2$ cathode exhibits a specific capacity of about 91 mAhg$^{-1}$ at a current density of 6 mAg$^{-1}$, whereas Na$_{0.74}$CoO$_2$ exhibits comparatively low specific capacity (70 mAhg$^{-1}$ at a current density of 6 mAg$^{-1}$). The cyclic voltammetry (CV) and electron impedance spectroscopy (EIS) were performed to determine the diffusion coefficient of Na, which found to be in the range of $10^{-10}$ cm$^2$s$^{-1}$.

## INTRODUCTION

Sustainable energy alternatives have become a global need in order to satisfy growing energy demands in recent years. Although recent developments in Li-ion battery (LIB) technology have created a benchmark in the energy requirement for large scale energy storage systems, the inadequacy of Li- ion in the earth's crust and high cost have led to a search for low cost battery system. In this direction, sodium-ion batteries (SIBs) have been considered as a promising alternative for energy storage devices owing to the large abundance and low cost of sodium [1]. However, larger ionic size of Na-ion (1.02Å) than Li-ion (0.76Å), is a major problem to explore efficient electrodes for high rate capability SIBs. Therefore, developing an efficient cathode material with high specific capacity and long cycle life is a bottleneck process. Layered metal oxides are considered as one of the highly reversible cathode materials due to their stable cycling performances, rich chemistry and capability to tolerate the high stress originated due to structural changes. These layered oxides are classified into two main groups, O3 type and P2 type, where "O" and "P" refers to the Na coordination (P-Prismatic, O-Octahedral) and "2" or "3" to the number of MO$_2$(M-transition metals) slabs in that particular cell [1]. In case of P2-Na$_x$CoO$_2$, there are two distinct prismatic sites; one is sharing only faces whereas the second one is sharing only edges with CoO$_6$ octahedra [2]. Usui *et al*. proposed Nb-doped rutile TiO$_2$ as an anode for SIBs, where Nb-doping significantly boosts the Na-ion storage performance due to the enhanced conductivity and broadened ionic channels [3]. These effects of Nb-doping are also expected for Na$_{0.74}$CoO$_2$ cathode material, which has never been reported to the best of our knowledge.

Here, we have prepared P2-Na$_{0.74}$Co$_{1-x}$Nb$_x$O$_2$ ($x$=0, 0.05) by a facile solid-state route. We observe improvement in the specific capacity and diffusion coefficient, which can be due to the widened Na-ion migration channels and much reduced charge transfer resistance with Nb substitution in Na$_{0.74}$CoO$_2$.

## EXPERIMENTAL DETAILS

We have synthesized polycrystalline Na$_{0.74}$CoO$_2$ (NCO) and Na$_{0.74}$Co$_{0.95}$Nb$_{0.05}$O$_2$ (NCNO) through solid-state reaction using precursors Na$_2$CO$_3$ (sodium carbonate), Co$_3$O$_4$ (cobalt oxide) and Nb$_2$O$_5$ (niobium pentoxide, for $x$=0.05 sample). Initially, these two materials with proper stoichiometric amount were grinded for 6-7 hours and then sintered directly at 800°C in a preheated furnace to obtain the final product. The crystal structure of the samples were obtained by x-ray diffraction (XRD) using CuKα radiation (λ =1.5406 Å) from Panalytical x-ray diffractometer in the 2θ range of 10–70°. The electrode was fabricated by casting the slurry on bare aluminium foil. We used doctor blade method to prepare the slurry, where active material, carbon and binder (Polyvinylidene fluoride, PVDF) are mixed with a initial weight ratio 8:1:1. The 2016 type coin-cells were assembled using a crimping machine (MTI Corp.) inside a nitrogen filled glove box (Jacomex). The electrolyte used was 1M NaClO$_4$ dissolved as the solute in 1:1ethylene carbonate (EC) and diethyl carbonate (DEC). We have used potentiostat (Palmsens) for CV, battery cycler (Bio-Logic-VMP3) for EIS and galvanostatic charging discharging using cycler from Neware.

## RESULTS AND DISCUSSION

The Rietveld refinement of room temperature XRD patterns of both the samples confirm the hexagonal crystal structure with space group: P63/mmc (no. 194), as shown in Figs. 1(a) and 1(b). We found the lattice parameters of $a=b=2.823$Å and $c=10.933$Å for NCO sample and $a=b=2.825$Å and $c=10.954$Å for NCNO sample. The convergence factor ($\chi^2$) is about 1.4, which confirms the good quality of fitting. The XRD results show that the Nb substitution slightly enhances the volume of the unit cell. The scanning electron microscopy (SEM) images in Figs. 1(c) and (d) reveal the platelet like morphology and confirm that the particles are typically in the μm range.

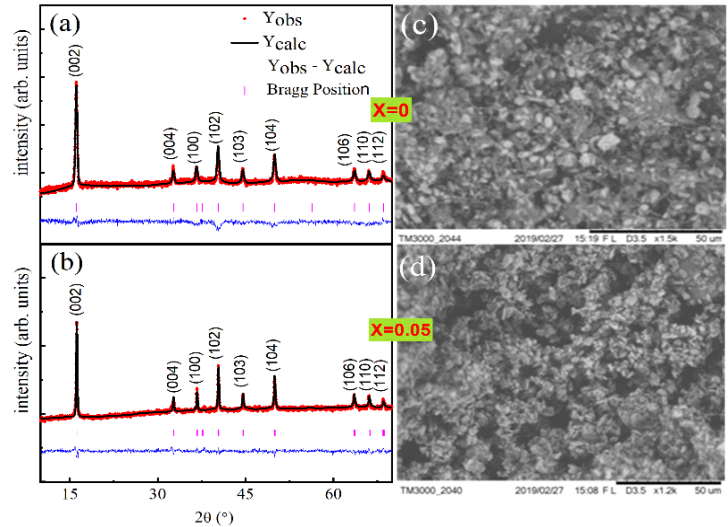

**FIGURE 1:** Room temperature XRD pattern (red) and Rietveld refinement (black) of as-prepared (a) NCO ($x=0$) and (b) NCNO ($x=0.05$), and SEM images of the (c) NCO (d) NCNO samples.

Figures 2(a) and 2(b) depict the Nyquist plots of freshly fabricated Na-NCO and Na-NCNO half cells, respectively where Na metal is used as counter electrode. Based on the nature of the Nyquist plots, we here design equivalent circuits for the EIS spectra, as shown in the insets of Figs. 2(a) and 2(b). The R1 represents the sum of Ohmic resistances (resistances of working electrode, electrolyte and current collector). In Fig. 2(a) inset Q1//R2 and Q2//R3 correspond to the capacitance and resistance in the high frequency (HF) and low frequency (LF) regions, respectively. Here, constant phase element (Q) is used instead of capacitor to show the non-ideal behaviour of the electrode due to rough/porous surface [4]. The straight line in the LF region is fitted by using a constant phase element, Q3 instead of warburg impedance. Warburg impedance is used to show the diffusion of the ions within the cathode. In Fig. 2(b), R2 is the charge transfer resistance, which reflects the direct charge transfer to and from the electrode surface and R3 refers the charge transfer resistance at the electrolyte/immobilised substances [5]. We can calculate the diffusion coefficient D from the LF region data of EIS using the following equation:

$$D = \frac{1}{2}\left[\frac{V_M}{FA\sigma}\right]^2 \left[\frac{dE}{dx}\right]^2 \qquad (1)$$

where, $V_M$ is the molar volume of the active material, $A$ is the area of the electrode, $dE/dx$ is the slope of the cell voltage to the Na ion concentration, which is procured from the plot in Fig. 2(i), $F$ is the Faraday constant (96486 C

mol$^{-1}$), σ (Warburg impedance coefficient) is the slope of linear fit between Z' and $\omega^{-0.5}$ because Z' and $\omega^{-0.5}$ depends linearly on each other [6]. Using equation (1), the diffusion coefficient values of NCO and NCNO are found to be $1.9 \times 10^{-10}$ and $9.3 \times 10^{-10}$ cm$^2$s$^{-1}$, respectively. Figures 2(c) and (d) show the CV curves of NCO and NCNO at different scan rates in the voltage range 2.0 to 4.0 V vs. Na/Na$^+$.

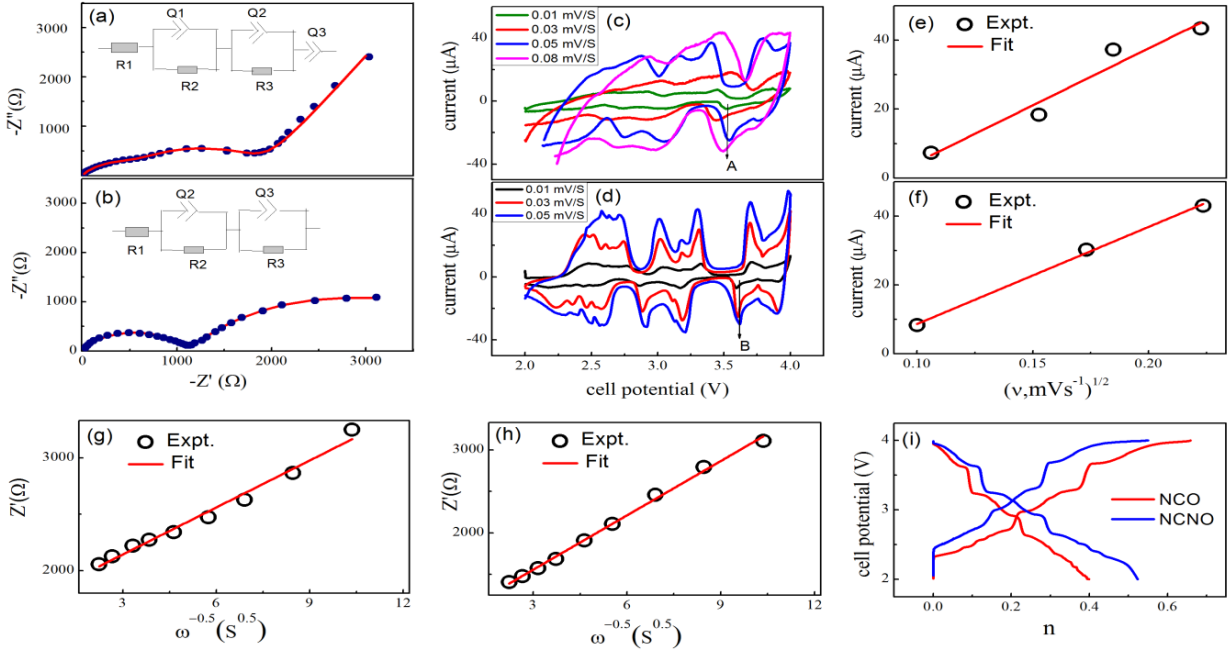

**FIGURE 2:** Nyquist plots of freshly fabricated (a) Na-NCO (b) Na- NCNO half cells, the equivalent circuit for analysis of the Nyquist plots are given as insets in corresponding panels, the cyclic voltammogram (between 2.0 and 4.0 V) of (c) NCO (d) NCNO electrodes at different scan rates, the relationship of the peak current ($i_p$) and the square root of scan rate ($v^{0.5}$) for (e) NCO (f) NCNO, linear fit of Z' and $\omega^{-0.5}$ (g) NCO (h) NCNO, (i) the variation of Na-ion concentration (n) during charge discharge at slow current density 6 mAg$^{-1}$.

The multiple pairs of redox peaks indicate the complex phase transition reactions in both the cells. Figures 2(e) and 2(f) show the linear dependence of normalised peak currents ($i_p$) and square root of scan rates ($v^{0.5}$). This confirms the relationship, showing reversible characteristic of the redox system [7]. If the system is diffusion controlled and the charge transfer rate at the interface is fast enough, then Randles Sevcik equation holds good for the reversible redox reaction [8]. This equation shows the relationship between the peak current and CV scan rate, which is:

$$i_p = (2.69 \times 10^5) A D_{Na^+}^{1/2} C n^{3/2} v^{1/2} \qquad (2)$$

where $i_p$ is the peak current, $n$ is the stoichiometric number of electrons involved in the electrode reaction (here $n$=0.1 for NCNO and 0.09 for NCO), $A$ is the active surface area involved in the reaction, $C$ is the concentration of Na$^+$ in the cathode, $D$ is the apparent diffusion constant of Na$^+$ and $v$ is the scan rate. From the above equation the apparent diffusion coefficient values of NCO and NCNO for the cathodic peaks (A and B) are calculated to be $1.66 \times 10^{-10}$ and $2.5 \times 10^{-10}$ cm$^2$s$^{-1}$, respectively.

The galvanostatic charging-discharging plateaus indicate the formation of different phases due to the intercalation/deintercalation of Na$^+$ ions. Figures 3(a) and 3(b) depict the galvanostatic charging-discharging curves, where specific capacities of 70 and 91 mAhg$^{-1}$ are obtained for NCO and NCNO, respectively, at current density 6 mAg$^{-1}$ (here 1C= 175 mAg$^{-1}$ for NCO and 172 mAg$^{-1}$ for NCNO). Figs. 3 (c) and (d) show the capacity retention of NCO and NCNO after 50 cycles. There is degradation in capacity from 70 mAhg$^{-1}$ to 53 mAhg$^{-1}$ for NCO and 91 mAhg$^{-1}$ to 80 mAhg$^{-1}$ for NCNO at 6 mAg$^{-1}$. This occurs because of the Nb substitution, which increases the inter-slab distance, providing larger space for Na-ion intercalation than NCO. Figures 3(e) and (f) depict the Coulombic efficiencies of about 100% for both NCO and NCNO samples at a current density of 9 mAg$^{-1}$ after 10-20 cycles.

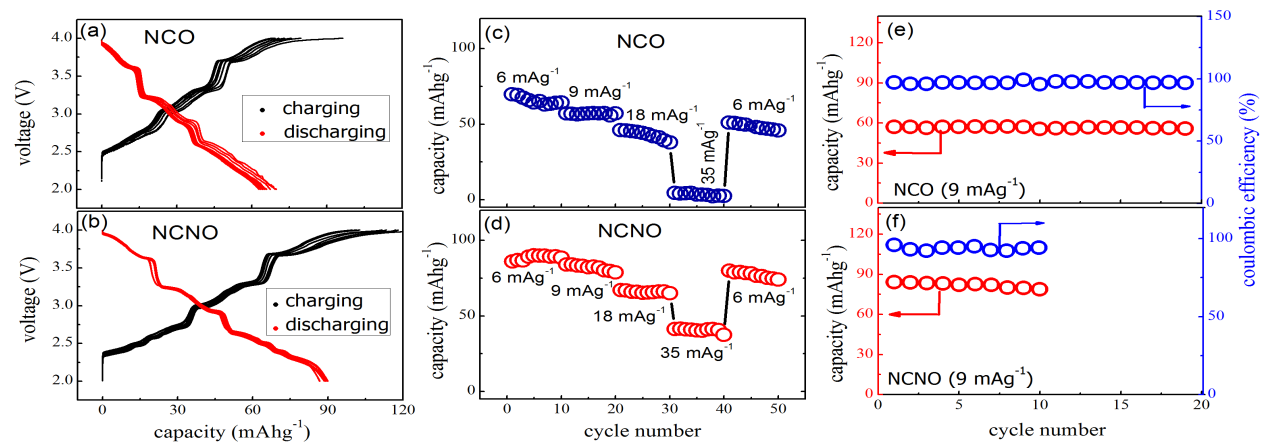

**FIGURE 3:** Galvanostatic charging-discharging at current density 6 mAg$^{-1}$ for (a) NCO (b) NCNO, specific capacity versus cycle numbers at different current densities for (c) NCO (d) NCNO, specific capacity and Coulombic efficiency change with cycle numbers at 9 mAg$^{-1}$ for (e) NCO (f) NCNO cells.

The above discussion shows the better cycle life of NCNO cathode than NCO. From the below figures we can conclude that substitution of Nb in Co site increases both the rate capability and cycle life significantly.

## CONCLUSIONS

In conclusion, we have successfully prepared NCO and NCNO by solid-state route and demonstrated as cathode materials for Na-ion batteries. The Rietveld refinement of XRD patterns confirms the hexagonal crystal symmetry. The SEM images show a platelet like morphology of the micrometer range particles. The specific capacities of about 70 mAhg$^{-1}$ and 91 mAhg$^{-1}$ at current density of 6 mAg$^{-1}$ have been obtained for Na–NCO and Na–NCNO cells, respectively, which show that NCNO has higher specific capacity and stable cycle life than NCO. The obtained diffusion coefficient values using CV and EIS measurements are in the range of $10^{-10}$ cm$^2$s$^{-1}$.

## ACKNOWLEDGEMENTS

This work is financially supported by SERB-DST through Early Career Research (ECR) Award to RSD (project reference no. ECR/2015/000159). JP and MC thank UGC and SERB-DST (NPDF, no PDF/2016/003565), respectively for providing fellowship. Authors acknowledge the physics department and central research facility (CRF) of IIT Delhi for providing XRD and SEM. We are thankful to Prof. Amit Gupta for providing necessary facilities to prepare coin cells and for EIS measurements.